%% file: nsdi25.tex
\documentclass[sigplan,10pt]{acmart}
\acmSubmissionID{1679}

\usepackage[english]{babel}
\usepackage{blindtext}
\usepackage{scalefnt}

\copyrightyear{2026}
\acmYear{2026}
\setcopyright{cc}
\setcctype{by}
\acmConference[EUROSYS '26]{21st European Conference on Computer Systems}{April 27--30, 2026}{Edinburgh, Scotland, UK}
\acmBooktitle{21st European Conference on Computer Systems (EUROSYS '26), April 27--30, 2026, Edinburgh, Scotland, UK}
\acmPrice{}
\acmDOI{10.1145/3767295.3769387}
\acmISBN{979-8-4007-2212-7/26/04}

\usepackage{tikz}
\usepackage{amsmath}
\usepackage{listings}
\usepackage{xspace}
\usepackage{colortbl}
\usepackage[group-separator={,},group-minimum-digits=4]{siunitx}
\usepackage{tabularx}
\usepackage{enumitem}
\usepackage{multirow}
\usepackage{graphicx, subcaption}
\usepackage{algorithm}
\usepackage{algpseudocode}
\usepackage{bold-extra}
\usepackage{balance}

\newcommand*{\mycommentstyle}[1]{%
  \begingroup
    \fontseries{lc}%
    \fontshape{it}%
    \selectfont
    \footnotesize
    \color{gray}
    \lstset{columns=fullflexible}%
    #1%
  \endgroup
}

\lstdefinestyle{python}{
    backgroundcolor=\color{white},
    commentstyle=\mycommentstyle,
    keywordstyle=\color{blue},
    numberstyle=\footnotesize\color{black},
    stringstyle=\color{purple},
    basicstyle=\ttfamily\footnotesize,
    breakatwhitespace=false,
    breaklines=true,
    captionpos=b,
    keepspaces=true,
    numbers=left,
    numbersep=5pt,
    showspaces=false,
    showstringspaces=false,
    showtabs=false,
    tabsize=2
}

\lstdefinestyle{cpp}{
    backgroundcolor=\color{white},
    commentstyle=\mycommentstyle,
    keywordstyle=\color{blue},
    numberstyle=\footnotesize\color{black},
    stringstyle=\color{purple},
    basicstyle=\ttfamily\footnotesize,
    breakatwhitespace=false,
    breaklines=true,
    captionpos=b,
    keepspaces=true,
    numbers=left,
    numbersep=5pt,
    showspaces=false,
    showstringspaces=false,
    showtabs=false,
    tabsize=2,
    morekeywords={m256i,size_t}
}
\usepackage{enumitem}

\newcommand{\sys}{\textsc{Fixpoint}\xspace}
\newcommand{\lang}{\textsc{Fix}\xspace}
\newcommand{\posixlib}{Flatware\xspace}

\usepackage{microtype}

\begin{document}
%-------------------------------------------------------------------------------

%don't want date printed
\date{}

\title{\lang: externalizing network I/O in serverless computing}

\author{Yuhan Deng}
\affiliation{%
  \institution{Stanford University}%
  \city{}
  \country{}%
}

\author{Akshay Srivatsan}
\affiliation{%
  \institution{Stanford University}%
  \city{}
  \country{}%
}

\author{Sebastian Ingino}
\affiliation{%
  \institution{Stanford University}%
  \city{}
  \country{}%
}

\author{Francis Chua}
\affiliation{%
  \institution{Stanford University}%
  \city{}
  \country{}%
}

\author{Yasmine Mitchell}
\affiliation{%
  \institution{Stanford University}%
  \city{}
  \country{}%
}

\author{Matthew Vilaysack}
\affiliation{%
  \institution{Stanford University}%
  \city{}
  \country{}%
}

\author{Keith Winstein}
\affiliation{%
  \institution{Stanford University}%
  \city{}
  \country{}%
}

%\author{\scalefont{0.835}{\mbox{Yuhan Deng, Akshay Srivatsan, Sebastian Ingino, Francis Chua, Yasmine Mitchell, Matthew~Vilaysack, Keith Winstein}}}
%\affiliation{%
%  \institution{\scalefont{0.835} Stanford University}%
%  \city{}
%  \country{}%
%}
\renewcommand{\shortauthors}{Deng, Srivatsan, Ingino, Chua, Mitchell, Vilaysack, and Winstein}

%-------------------------------------------------------------------------------
\begin{abstract}
%-------------------------------------------------------------------------------
  We describe a system for serverless computing where users, programs,
  and the underlying platform share a common representation of a
  computation: a deterministic procedure, run in an environment
  of well-specified data or the outputs of other computations. This
  representation externalizes I/O: data movement over the network is
  performed exclusively by the platform. Applications can describe the
  precise data needed at each stage, helping the provider schedule
  tasks and network transfers to reduce starvation. The design
  suggests an end-to-end argument for outsourced computing, shifting
  the service model from ``pay-for-effort'' to ``pay-for-results.''
\end{abstract}




\maketitle

\input{introduction}
\input{relatedwork}
\input{fix}

\input{implementation}
\input{evaluation}

\input{limits}

\input{concl}

\input{acks}

\balance
% \input{artifact}

%-------------------------------------------------------------------------------
%\section*{Acknowledgments}
%-------------------------------------------------------------------------------
%
%The USENIX latex style is old and very tired, which is why
%there's no \textbackslash{}acks command for you to use when
%acknowledging. Sorry.

%-------------------------------------------------------------------------------
%\section*{Availability}
%-------------------------------------------------------------------------------

%USENIX program committees give extra points to submissions that are
%backed by artifacts that are publicly available. If you made your code
%or data available, it's worth mentioning this fact in a dedicated
%section.

%-------------------------------------------------------------------------------
\bibliographystyle{plain}
\bibliography{\jobname}

%%%%%%%%%%%%%%%%%%%%%%%%%%%%%%%%%%%%%%%%%%%%%%%%%%%%%%%%%%%%%%%%%%%%%%%%%%%%%%%%
\end{document}

%% file: introduction.tex
\raggedbottom
\section{Introduction}

For a decade, cloud-computing operators have offered ``serverless''
function-as-a-service products. These systems let users upload
functions to be invoked on request. When this happens, the function is
allocated a slice of a physical machine's RAM, CPU, and NIC, and the
customer is billed for the time until it finishes~\cite{awslambda,
azurefunction}. In practice, cloud functions are typically used for
asynchronous services where each invocation runs independently, but
researchers have also explored their use for large jobs that launch
thousands of parallel invocations working together with complex
dataflow: video processing~\cite{excamera}, linear
algebra~\cite{pywren, numpywren}, software compilation and
testing~\cite{gg}, theorem proving~\cite{wu2020parallelization}, 3D
rendering~\cite{r2e2}, ML training~\cite{lambdaml}, data
analysis~\cite{flint}, sorting~\cite{flashburst}, etc.

Despite this interest, effective use of serverless computing remains
elusive. In this paper, we argue that a root cause is an
\emph{underconstrained notion of networked computation}, one where the I/O and dependencies
of user functions are opaque to the platform. Consider a common
serverless application: a cloud function that resizes an
image~\cite{lambda-thumbnail}. A user creates the function by
uploading a piece of code---call this $f$. When the function is
invoked, the provider finds a physical server with enough available
RAM and cores, transfers and unpacks the code if not already present,
claims a slice of RAM, and runs the function, generally as a Linux
process in a pre-warmed VM. After seeing the invocation payload (an
HTTP request or other event), the function requests the
image file~$x$ from network storage, e.g.~Amazon~S3.

From the user's point of view, the invocation was always meant to
compute $f(x)$ (the resized image), but from the provider's
perspective, $f$ is a running Linux process, and its dependency on $x$
wasn't known until after the code was placed and running. If S3 has
cached $x$ nearby, the retrieval happens quickly. Otherwise, the
function will wait, occupying and mostly idling its slice of RAM until retrieval
finishes.% (RAM oversubscription and paging to disk are
%generally infeasible in this context of short
%functions~\cite{firecrackerballoon}.)

For computations that are short relative to network and storage
  latencies~\cite{granular, flashburst}, limitations of this service
  model can be significant. If the user had been able to express that
  the invocation represented ``$f(x)$'' in a way the provider
  understood, the provider might have attempted a better strategy to place
  or schedule it, e.g.:
  {
  \small
\renewcommand\labelitemi{\raisebox{0.07 \baselineskip}{\tiny$\bullet$}}
\begin{itemize}[topsep=0pt, itemsep=0pt, leftmargin=3.5ex]
\item simultaneously transfer $f$ (the code) and $x$ (the image) to the server so the task can finish sooner,
\item wait to allocate $f$'s RAM and CPU until $x$ arrives,
letting other functions run there in the meantime,
\item instead of starting $f$ on a server chosen without regard to $x$, choose a server close to whichever dependency ($f$ or $x$) is bigger,
\item delay the invocation, hoping to aggregate multiple tasks that depend on $x$ to run in a batch on one server,
\item if $f(x)$ is part of a pipeline $g(f(x))$, and if $y = f(x)$ is hinted to be large, then transfer $f$ and $x$ close to their \emph{downstream} dependency~$g$, and run $f(x)$ on that server before running $g$ on the result---avoiding the need to transfer $y$ over the network,
\item or if $x$ can be computed deterministically by $h(z)$, then if easier, fetch $h$ and $z$ and recompute~$x$ instead of transferring it.
\end{itemize}
}

In many cases, these strategies could improve the job throughputs,
latencies, RAM and CPU utilization, and perhaps costs of serverless
platforms. But they probably aren't feasible today, even for an
image-resizing function of one input, because $f$'s dataflow was
``internal'': it fetched~$x$ by opening a socket, sending a request,
and receiving arbitrary data. Even if $x$ came from the provider's own
storage service, the provider didn't observe the dependency until
after the task was placed and running. For the sorts of jobs surveyed
in the first paragraph, jobs that launch thousands of parallel invocations
with complex dataflow among them, the need for good placement and
scheduling will be even greater.% Moreover, providers are not strongly
%incentivized to make jobs complete faster when users pay for each
%millisecond of runtime.

This paper presents \lang, an architecture for serverless computing that
\emph{externalizes} I/O, making application dataflow visible to, and
performed by, the underlying networked system. In \lang, function
invocations are described in a low-level ABI (application binary
interface) that specifies a sealed container where execution occurs,
containing dependencies that are addressed in a way the program and provider
both understand---maybe as the output of another invocation.

In \lang, programs can choose to capture only the minimum data needed to
make progress at each step of a larger job. The underlying platform
uses its visibility and flexibility to place and schedule tasks
and transfers to reduce starvation and use of the network, e.g.~via
the strategies above.

This paper's main contribution is in \lang's design and the
demonstration that I/O externalization, with the ability to express
precise and dynamic data-dependencies with little overhead, can boost
performance and efficiency. \lang is a realization of I/O-compute
separation~\cite{monotasks, earnest, dandelion, dandelion2,
klimovicrethinking} as well a mechanism for programs to provide the
platform with visibility---perhaps partial visibility, refined as
computation proceeds---into future data- and control flow. \lang does
this in a declarative way that can be parsed anywhere, avoiding
round-trips to a scheduler when invoking a new task.

\lang's design and implementation have a number of mutually
reinforcing characteristics that lead to efficient execution. \lang's
invocations are concisely described in a packed binary format designed
to minimize runtime overhead.  We implemented a runtime for
\lang, called \sys, that has a per-invocation overhead of about
\qty{1.5}{\micro\second}. This means that applications can afford to
use fine-grained containers that capture only data
needed to make progress at each stage. Minimizing the data
``footprint'' of each invocation helps \sys reduce cold-start times
and optimize the scheduling and utilization of CPUs, RAM, and the network.
% \lang lets applications express partial knowledge of downstream
%dataflow beyond an individual invocation, which \sys uses to optimize
%performance.

\lang has significant limitations. It represents a
constrained model of computation: to describe each task in a
placement-agnostic way, invocations must be of pure functions applied
to content-addressed data or to the outputs of other
invocations. Functions can't access data outside the container. At least at
present, \lang doesn't support calls to nondeterministic services
e.g., clocks, true random number generators, multi-user databases, or
arbitrary Web APIs. \lang is its own ABI and doesn't run Linux
executables; it runs some POSIX programs (e.g.~CPython, clang) but we
had to recompile them with a \lang-targeting toolchain to achieve
this.  We haven't measured \lang's ease of use or effect on developer productivity.

\vspace{0.5 \baselineskip}

\noindent \textbf{Summary of results.} We found that \lang's approach can unlock significant advantages in performance and efficiency
(as well as reproducibility and reliability, aspects we did not
evaluate quantitatively). We evaluated several applications run on
\sys, compared with OpenWhisk, MinIO, and Kubernetes (open-source
analogs of AWS Lambda and S3), Pheromone~\cite{pheromone}, and
Ray~\cite{ray}; full results are in Section~\ref{sec:eval}.

\sys creates hermetic containers without spawning
OS processes, by requiring that functions be
converted ahead-of-time to safe machine code. This results in lower
overhead than systems based on Linux containers (OpenWhisk) or
higher-level programming languages (Ray). To invoke a trivial function
that adds two 8-bit integers, \sys's containers show lower overhead
(fig.~\ref{fig:lwvirt}):

\vspace{0.5\baselineskip}

{
\small

\noindent \begin{tabularx}{\columnwidth}{l r l}
    Approach & Time & slowdown vs.~\lang \\
    \hline
    \lang & \qty{1.46}{\micro\second} & 1$\times$ \\
    Linux \footnotesize \texttt{vfork}+\texttt{exec} & \qty{449}{\micro\second} & 307$\times$ \\
    Pheromone & \qty{1.05}{\milli\second} & 720$\times$ \\
    Ray & \qty{1.29}{\milli\second} & 881$\times$ \\
    Faasm & \qty{10.6}{\milli\second} & 7,260$\times$ \\
    OpenWhisk & \qty{30.7}{\milli\second} & 20,980$\times$ \\
  \end{tabularx}

}

\vspace{0.5\baselineskip}

In a different experiment, we used Linux's CPU-state statistics to
measure how much of these gains come from avoiding starvation---by
co-scheduling computations and transfers, and waiting to
allocate CPU and RAM until dependencies have arrived. We wrote a
program to count non-overlapping strings in a 96~GiB
dataset from Wikipedia and ran it on a 320-core, 10-node
cluster. \lang's approach avoids a substantial amount of CPU
starvation (fig.~\ref{fig:wikipedia}):

\vspace{0.5\baselineskip}

{
\small

\noindent \begin{tabularx}{\columnwidth}{l r r}
  Approach & Time & CPU waiting \% \vspace{-3 pt} \\
  & & \scriptsize ({\tt idle} + {\tt iowait} + {\tt irq}) \\
    \hline
    \lang & \qty{3.25}{\second} & 37\% \\
    \lang \scriptsize (with ``internal'' I/O) & \qty{33.8}{\second} & 92\% \\
    OpenWhisk + MinIO + K8s & \qty{63.9}{\second} & 92\% \\
  \end{tabularx}
}

\vspace{0.5\baselineskip}

Finally, we implemented a key-value store represented on disk as a
B+-tree, using \lang and two other approaches. Each version traverses
the B+-tree node-by-node to retrieve the value corresponding to a
key. As we decrease the maximum number of children of each B+-tree
node, this process results in a smaller memory footprint and total
amount of data accessed, at the cost of more function
invocations. Compared with Ray, \lang's semantics let users benefit
from breaking down programs with fine granularity
(fig.~\ref{fig:bptree}):

\vspace{0.5\baselineskip}

{
\small

\noindent \begin{tabularx}{\columnwidth}{l r l}
    Approach \footnotesize (B+-tree of arity 256) & Time & \footnotesize slowdown vs.~\lang \\
    \hline
    \lang & \qty{0.14}{\second} & 1$\times$ \\
    Ray & \qty{2.8}{\second} & 19.6$\times$ \\
    Ray \scriptsize (broken into fine-grained invocations) & \qty{5.74}{\second} &
    40$\times$ \\
  \end{tabularx}

}

\vspace{0.5\baselineskip}

\lang represents a fundamentally different approach to
outsourced computing: one that's more constraining and probably more
difficult to program for, but ultimately advantageous for customers
(whose jobs run faster) and providers (whose infrastructure is used
more efficiently). Current service abstractions represent something of
a ``pay-for-effort'' system---by billing customers for each
millisecond that a function occupies a machine slice, idle or not,
providers aren't directly incentivized to improve scheduling and
placement. Even if a provider wanted to do this, current systems lack
the visibility into application dataflow to do it well. \lang's
approach suggests a shift towards ``pay-for-results'': computations
described in a way that permits providers to innovate in the placement
and scheduling of computation and I/O, so long as they arrive at the
correct answers.

This paper proceeds as follows. In section~\ref{sec:relwork}, we
discuss the substantial context of related work across several areas. We
describe \lang's design (sec.~\ref{sec:design}) and its implementation
in the \sys runtime (sec.~\ref{sec:impl}). We report our evaluation
in section~\ref{sec:eval}, finishing with limitations
(sec.~\ref{sec:limits}) and a conclusion.

%% file: relatedwork.tex
% should talk about Adam Belay's recent work

% also "what about just prefetching hints?"

\section{Related work}
\label{sec:relwork}

\lang relates to prior work across workflow orchestration
(Hadoop ~\cite{hadoop}, Spark~\cite{spark}, etc.), techniques that optimize
serverless platforms with lightweight containers for dense packability or
locality hints, tools that run highly parallel workloads on current
function-as-a-service platforms, containerization and execution systems
(Docker~\cite{docker}, NixOS~\cite{nixos}, etc.), and content-addressed
storage. We discuss how \lang relates to this prior literature in several
areas.

\textbf{Cluster orchestration systems. } Cluster orchestration systems
like Spark~\cite{spark}, Dryad~\cite{dryad}, CIEL~\cite{ciel} and
Ray~\cite{ray} allow programmers to express applications as a group of
tasks, and orchestrate execution of the tasks across a cluster. Task
interdependencies can be represented at runtime as a static DAG (Spark
and Dryad) or dynamic task graph (CIEL and Ray). These systems
generally employ language-level mechanisms: users spawn tasks using
domain-specific languages (CIEL and Dryad), or with a
pre-existing programming language (Python for Ray, Scala for Spark,
etc.)

\lang's computation model represents interdependencies in a
dynamic graph (similar to CIEL or Ray), in a somewhat more general
sense: \lang's invocations describe all data-dependencies that code
will have access to; \lang can capture subselections of existing data objects
and the relationships between application data structures (e.g.~the
relationships between nodes in a
B+-tree); this kind of dataflow can't generally be exposed to the
runtime by current systems.

\lang enforces I/O externalization: all data-dependencies that code need access
to must be made explicit. This allows it to freely schedule tasks at different
execution locations. In comparison, existing systems allow programs to states
not captured by the computation representation, such as local filesystem. This
makes \lang more amenable to outsourcing computations to cloud services.

Previous work relies on runtime infrastructure to track dependency
information at a centralized scheduler or a designated physical
node. In contrast, \lang unifies the description of data flow---inputs
and outputs of invocations---with control flow---which function should
be invoked with the results of another---in a single serializable
format. Dependency information is shipped with data defining a
function, avoiding round-trips.  This leads to lower
dependency-resolution overhead that allows finer-grained function
invocations.

\textbf{Scheduling and containers for serverless platforms.} Much
prior work is aimed at optimizing the performance of
function-as-a-service platforms with conventional architectures for applications
with interdependent workflows. This includes adding long-living caches beyond
individual function invocations~\cite{romero2021faa}, and providing locality
hints~\cite{palette} to for better placement decisions. The line of work most
similar to \lang is workflow-based serverless systems~\cite{step-function,
pheromone} with a static function dependency model, e.g. the outputs
of a function $f$ are always consumed by another function $g$. \lang represents
data-dependencies in a richer way at a finer-grained per-invocation level. Prior
work has proposed the model of I/O-compute separation~\cite{earnest} and
realization of the model~\cite{dandelion,dandelion2} that targets at better
elasticity for spiky serverless workloads. \lang focus on designing the
abstraction and mechanism for representing computational workloads in a
I/O-compute separation way.

Another line of work designs lightweight containers to allow denser packability, such as
Firecracker~\cite{firecracker}, Virtine~\cite{virtine}, AlloyStack~\cite{alloystack},
Junction~\cite{junction}, Faasm~\cite{faasm}, and WasmBoxC/wasm2c~\cite{wasmboxc}. As
part of our work on \lang, we became significant contributors to and maintainers
of the wasm2c codebase; \lang's toolchain includes this tool.

\textbf{Massively burst-parallel applications.} There has been
considerable interest in using serverless platforms for short-lived,
large-scale, highly parallel jobs, including video
processing~\cite{excamera}, linear algebra~\cite{pywren, numpywren},
software compilation and testing~\cite{gg}, theorem
proving~\cite{wu2020parallelization}, 3D rendering~\cite{r2e2}, ML
training~\cite{lambdaml}, data analysis~\cite{flint, wukong}, etc.
\lang aims to be a better platform for these kinds of applications.

\textbf{Build environments and content-addressed
  storage. } \lang's computation-addressed
dependencies for user programs resemble execution-environment
languages like Docker~\cite{docker}, NixOS~\cite{nixos}, or
Spark~\cite{spark} (discussed above). \lang's binary representation of
dependencies draws inspiration from content-addressed systems such as
Git~\cite{git}, Bittorrent~\cite{bittorrent}, Named Data
Networking~\cite{ndn}, and IPFS~\cite{ipfs}.

%% file: fix.tex
\section{\lang: describing dataflow to the runtime}

\label{sec:design}

In this section, we describe the design of \lang: a low-level binary
representation, or ABI, where code externalizes its data- and control
flow. Instead of fetching data by making network connections or
syscalls and waiting for a reply, programs describe the code and data
they need declaratively, in a format that's parsed and executed by the
runtime infrastructure.

\lang objects represent pieces of data, function invocations,
dependencies, and data sub-selection, in an in-memory representation
that is independent of programming language and placement on a
server. User functions make dataflow visible by constructing \lang
objects; a runtime can exchange references to these objects with
native functions using a defined calling convention. The computation
graph necessary to evaluate a \lang object is described by the
object itself, so runtimes do not need to maintain additional
metadata.

\lang's design is intended to let pieces of black-box machine code
precisely express their data needs in a manner lightweight enough to
permit microsecond-level overheads, but general enough to support
arbitrary applications, including ones where the dataflow graph
evolves over the course of a computation in a data-dependent way. To
enable efficient and flexible execution, \lang's design goals were:
\begin{enumerate}[noitemsep,topsep=0pt,itemsep=0pt]
\item Code can be represented as black-box machine code that
  originated from any programming language.

\item The complete data ``footprint'' needed to evaluate a
  function call will be known before it is invoked.

\item A function will always run to completion without blocking, and
  will finish execution without invoking another function or enlarging
  its data ``footprint.''
  
\item Functions will have tools to subselect from large data structures
  to fetch only the portion truly needed.
\end{enumerate}

These considerations led to the design below.

\subsection{Data and References}

\lang gives user functions an interface for expressing their data- and
control flow. \lang models two core types of Data:

\begin{description}[noitemsep]
\item[Blob]   A region of memory (an array of bytes).
\item[Tree]   A collection of other \lang Handles.
\end{description}

\lang also provides four reference types as Handles, each of which has a
particular binary representation in the ABI:

\begin{description}[noitemsep]
  \item[Object] A reference to accessible Data.
  \item[Ref]    A reference to inaccessible Data.
  \item[Thunk]  A reference to a deferred computation.
  \item[Encode] A request to evaluate a Thunk, and replace it by the result.
\end{description}

\begin{figure*}[ht]
  \centering
  \includegraphics[width=\linewidth]{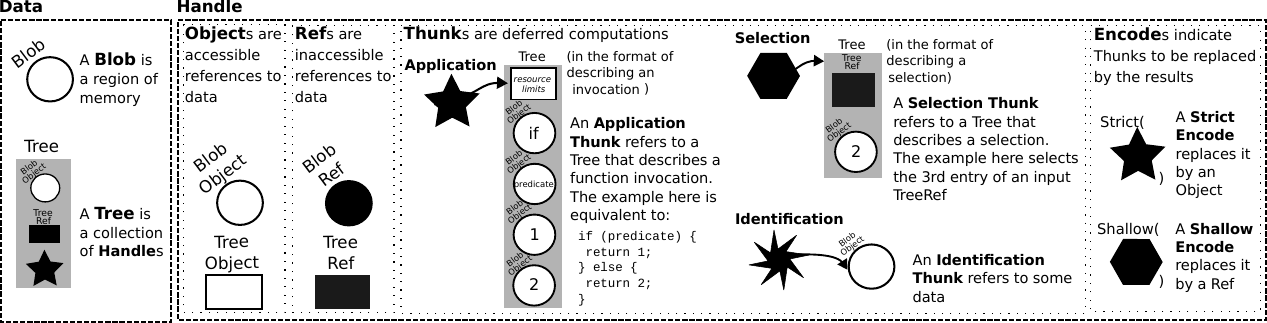}
  \caption{\lang represents function invocations and data dependencies with a
  unified serializable format.}
\end{figure*}

\subsection{Representation}

\paragraph{Handles.}

Every value in \lang is assigned a unique deterministic Handle.  These Handles 
contain information about a value's data, its type, and its size. This takes
the form of a truncated 192-bit BLAKE3 hash of the data, 16 bits of metadata
and type information, and a 48-bit size field. As an optimization, Blobs 30
bytes or smaller are stored as ``literals''---putting the Blob contents
directly in the Handle. Handles can be held and passed in machine SIMD registers,
e.g.~\texttt{\%ymm} on x86-64.

\paragraph{In-Memory Representation.}

Data are represented in an efficient format that minimizes copying;
Blobs are a contiguous sequence of bytes, and Trees are a
sequence of Handles.

\paragraph{Objects.}

An Object is a reference to accessible data directly needed for the
next function invocation. A function that is passed an Object may read its
data.

\paragraph{Refs.}

A Ref is a reference to inaccessible data not directly needed for the next
function invocation, but may be needed by a downstream consumer. A
function that is passed a Ref may look up the type and length of the
referent but not its data. Refs allow \lang functions to reference
remote data without fetching it to the execution server.

\paragraph{Thunks.}

A Thunk represents a function invocation whose value is not yet
needed, letting \lang functions refer to data which can be lazily
computed when needed. ``Application'' Thunks describe the execution of
a function in a container of available data and memory resource
limits.  There are two more styles of Thunk corresponding to common
usage patterns.  ``Identification'' Thunks represent the identity
function.  ``Selection'' Thunks represent a ``pinpoint''
data-dependency: the extraction of a \emph{subrange} of a Blob or a
Tree.

\paragraph{Encodes.}

Encodes\footnote{an \textbf{E}xplicit \textbf{N}amed
\textbf{C}omputation \textbf{o}n \textbf{D}ata or \textbf{E}ncodes}
embody a request to evaluate a particular Thunk.  When provided in the
input to a child function, they will be replaced with the result of
evaluating the Thunk.  There are two styles of Encode, Shallow and
Strict.

Shallow Encodes request the minimum amount of computation (or data
movement) needed to make meaningful progress. This means that a Thunk
is evaluated until the result is not a Thunk, and the result is
provided as a Ref. Strict Encodes request the maximum amount of
computation (or data movement) possible. A Thunk is replaced by its
fully-evaluated result as a Object, recursively descending into any Trees and
evaluating all Thunks within. These two styles of Encode allow \lang
programs to express both lazy and eager styles of evaluation.

\subsection{Minimum Repositories}

Each Thunk has the ability to access a bounded set of resources,
including both \lang data and hardware resources like RAM.  In order
to ensure that a function won't require any I/O operations before
finishing, the runtime must ensure these resources are available
throughout its execution. This set is called the ``minimum
repository'' of the Thunk.  While a function may not change its
minimum repository, it may create new Thunks with different minimum
repositories:

\begin{enumerate}[noitemsep,topsep=0pt,itemsep=0pt]
  \item It may specify new resource limits in the new Thunk;
  \item It may grow the repository by including an Encode, which will evaluate
    its referent and add the result to the new repository;
  \item It may shrink the repository by excluding data which were part of its
    own repository from the new Thunk.
\end{enumerate}

Since a function can't directly call another one (can't directly
evaluate a Thunk), these operations don't affect the minimum
repository of the currently-running function.

\subsection{Expressivity}

\lang enables programs to express their dataflow at fine granularity by
specifying the minimum repositories of child functions.  Programs are required
to express, at worst, an overapproximation of their data dependencies; it's
impossible for a program to access any data not explicitly requested.  \lang's
different types allow programs to describe complex access patterns to \sys.  A
pseudocode version of the \lang API is shown in Table~\ref{table:papi}; the
exact implementation varies depending on the implementation language.

\begin{figure}[ht]
\begin{minipage}{0.45\textwidth}
  \centering
  \footnotesize
  \begin{tabular} { l l }
  Function & Description \\
  \hline
    \texttt{T \textbf{read\_blob}(BlobObject)} & Read a Blob into a variable.\\
    \texttt{Value[] \textbf{read\_tree}(TreeObject)} & Read a Tree into an array. \\
    \texttt{BlobObject \textbf{create\_blob}(T)} & Create a Blob from a variable. \\
    \texttt{TreeObject \textbf{create\_tree}(Value[])} & Create a Tree from an array. \\
    \texttt{Thunk \textbf{application}(Tree)} & Apply a function (lazily). \\
    \texttt{Thunk \textbf{identification}(Value)} & Apply the identity function. \\
    \texttt{Thunk \textbf{selection}(Value, int)} & Select a child element. \\
    \texttt{Encode \textbf{strict}(Thunk)} & Strictly evaluate a Thunk. \\
    \texttt{Encode \textbf{shallow}(Thunk)} & Shallowly evaluate a Thunk. \\
  \hline
  \end{tabular}
\end{minipage}
\captionof{table}{\lang Pseudocode API}
\label{table:papi}
\end{figure}

\begin{figure}[h]
\begin{minipage}{\columnwidth}
\begin{minipage}{0.7\textwidth}
\begin{algorithm}[H]
\footnotesize
\hsize=\textwidth % <--------- THE HACK!
\caption{If Procedure}
\label{if}
\begin{algorithmic}
  \State $[rlimit, if, pred, a, b]$ $\gets$ \Call{read\_tree}{$input$}
  \If { \Call{read\_blob}{$pred$} }
    \State \Return $a$
  \Else
    \State \Return $b$
  \EndIf
\end{algorithmic}
\end{algorithm}
\end{minipage}\hfill
\begin{minipage}{0.25\textwidth}
    \centering
    \includegraphics[scale=1]{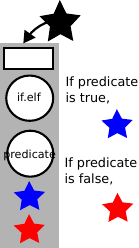}
\end{minipage}
  \captionof{figure}{This \texttt{if} procedure reads a boolean predicate from its input and selects one of two Thunks to return. }
  \label{fig:if}
\end{minipage}
\end{figure}

\lang's Thunks are lazy by default, which allows control flow to be expressed
by user-provided programs. For example, Fig.~\ref{fig:if} shows how a
user-provided \texttt{if} procedure can lazily select one of two Thunks based
on a predicate.  The other Thunk, and its data dependencies, never need to be
loaded or executed by \sys. The minimum repository of the \texttt{if} procedure includes
the machine codelet and the predicate, but excludes the Thunks'
definitions or results.

\begin{figure}[h]
\begin{minipage}{\columnwidth}
\begin{minipage}{0.65\textwidth}
\begin{algorithm}[H]
\footnotesize
\hsize=\textwidth % <--------- THE HACK!
\caption{Fibonacci Procedure}
\label{fib}
\begin{algorithmic}
    \State $[rlimit, fib, add, x] \gets$ \Call{read\_tree}{$input$}
    \If { $x = 0 \lor x = 1$ }
       \State \Return \Call{create\_blob}{$x$}
    \EndIf
    \State $x_1 \gets$ \Call{create\_blob}{$x - 1$}
    \State $t_1 \gets$ \Call{create\_tree}{[$rlimit$, $fib$, $add$, $x_1$]}
    \State $e_1 \gets$ \Call{strict(application}{$t_1$)}
    \State $x_2 \gets$ \Call{create\_blob}{$x - 2$}
    \State $t_2 \gets$ \Call{create\_tree}{[$rlimit$, $fib$, $add$, $x_2$]} 
    \State $e_2 \gets$ \Call{strict(application}{$t_2$)}
    \State $t_{sum} \gets$ \Call{create\_tree}{[$rlimit$, $add$, $e_1$, $e_2$]}
    \State \Return \Call{application}{$t_{sum}$}
\end{algorithmic}
\end{algorithm}
\end{minipage}
\begin{minipage}{0.32\textwidth}
    \centering
    \includegraphics[scale=0.8]{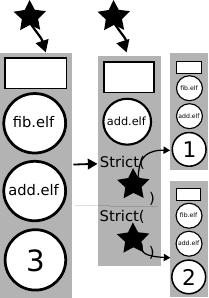}
    \vspace{0.1cm} 
\end{minipage}
  \captionof{figure}{
    A \lang version of the Fibonacci algorithm creates recursive Thunks and passes them to an addition procedure.
}
  \label{fig:fib}
\end{minipage}
\end{figure}

\begin{figure*}[ht]
\begin{minipage}{0.45\textwidth}
\begin{algorithm}[H]
\footnotesize
\hsize=\textwidth % <--------- THE HACK!
\caption{Get File Procedure}
\label{if}
\begin{algorithmic}
  \State $[rlimit, elf, path, info, dir] \gets$ \Call{read\_tree}{$input$}
  \State $i \gets$ index of child directory given $path$ and $info$
  \State $path_{new} \gets$ updated $path$
  \State $child \gets$ \Call{selection}{$dir$, $i$}
  \If { $path_{new} =$ ``'' } 
    \State \Return $child$
  \EndIf
  \State $info \gets$ \Call{selection}{$child$, $0$}
  \State $x_0 \gets$ \Call{strict}{$info$}
  \State $x_1 \gets$ \Call{shallow}{$child$}
  \State $res \gets$ \Call{create\_tree}{[$rlimit$, $elf$, $path_{new}$, $x_0$, $x_1$]}
  \State \Return \Call{application}{$res$}
\end{algorithmic}
\end{algorithm}
\end{minipage} \hfill
\begin{minipage}{0.52\textwidth}
    \centering
    \includegraphics[scale=0.85]{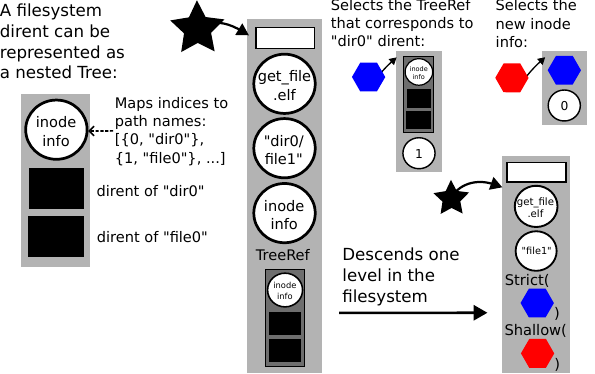}
\end{minipage}
  \captionof{figure}{
    A \lang procedure identifies the index of ``dir0'' within a TreeRef, and
    recurses into ``dir0''.  It specifies that the inode information from
    ``dir0'' is immediately necessary, while still providing a TreeRef to
    ``dir0'' for later use.
  \label{fig:selection}
}

\end{figure*}

By returning a Thunk, as in Fig.~\ref{fig:fib}, a function can
recurse or call other functions.  In this case, a user-provided
\texttt{fibonacci} procedure creates two Thunks corresponding to recursive
calls of itself, and a third Thunk adding the two results.  The
first two Thunks are wrapped in Strict Encodes, specifying that the addition
function needs the evaluated results.

Selection Thunks provide efficient access to large or
partially~evaluated data structures without including the entire structure in a
program's minimum repository.  Fig.~\ref{fig:selection} shows how a \lang
procedure may recursively descend a directory structure without fetching
the full contents of the directories or the files within.  The procedure uses a
strictly-encoded Selection Thunk to express its immediate dependency on the
inode information of the subdirectory, and a shallowly-encoded
Selection Thunk to express its eventual dependency on children of the
subdirectory.  In this way, the directory contents are never added to the
minimum repository of a Thunk, avoiding unnecessary data transfers and RAM usage.

%% file: implementation.tex
\section{\sys: a \lang runtime} \label{implementation}

\label{sec:impl}

In this section, we describe the design and implementation of \sys: a Linux application that functions
as a multi-node runtime for programs expressed in the \lang ABI.

\subsection{Procedures}

\sys procedures are implemented as machine codelets: small programs compiled to
native x86-64 machine code.  Each codelet contains an entrypoint,
\texttt{\_fix\_apply}, which is invoked when a Thunk
referring to that procedure is run.  This function receives as input the Handle
of a TreeObject (the Thunk's definition), and returns the Handle of a \lang object.
These procedures are provided as Executable and Linkable Format (ELF) files,
which are loaded and linked against the \sys API in a shared address space.
\sys requires these codelets to be sandboxed, but they may otherwise contain
arbitrary code.

\subsubsection{Safety and Correctness}

\sys runs arbitrary machine codelets within a single address space.
However, since these codelets are user-provided and therefore untrusted, \sys
must ensure they satisfy these properties.

%A particular execution of \sys defines a trusted \lang program which is
%responsible for generating safe, deterministic codelets.  This program's output
%should be a Tag which marks an ELF machine codelet as ``Runnable''.  When
%executing a Thunk, \sys will check that its procedure is indeed an ELF Blob
%contained within a ``Runnable'' Tag, produced by this trusted program.

A particular execution of \sys defines a trusted \lang program which is
responsible for generating safe codelets. The approach \sys currently uses is
to treat the trusted program as a trusted compilation toolchain for a
higher-level intermediate representation (IR) which provides sandboxing
capabilities. The current implementation uses Web\-Assembly (Wasm) \cite{wasm}
as its higher-level IR. Wasm provides memory-safety and determinism,
and is amenable to ahead-of-time compilation, at the cost of having a larger
trusted codebase than a verifier-based approach.
%As \sys only requires that the output binary be deterministic, variations in
%the compilation toolchain (e.g., new optimizations in a newer toolchain
%version) do not affect correctness; from \sys's perspective, this is a
%different program.

However, since \sys doesn't place any restrictions on what the \textit{input}
of this program should be, other approaches are possible; the use of
Wasm is not intrinsic to \lang or \sys.  For example, another approach
is to statically analyze machine code, as in Native Client~\cite{nacl64} or Deterministic Client~\cite{decl}. These approaches could also provide good performance and
memory-safety with a small trusted codebase.

While there are other representations that provide similar guarantees,
Wasm has the benefits of strong existing toolchain and language
support. This allows existing third-party software, such as the CPython
interpreter or clang compiler, to be easily ported to \lang. Wasm also
provides programs the ability to reference opaque pieces of external data (via
an \texttt{externref}), which \sys uses to efficiently exchange \lang Handles
with the otherwise-untrusted function invocations while maintaining security.

We have implemented an ahead-of-time ``trusted toolchain'' that takes
programs that have been compiled to Wasm and compiles them, in
turn, into x86-64 machine codelets.  Given a Wasm module, \sys
compiles it by using (1) the Wasm-spec-conforming \texttt{wasm2c} tool~\cite{wasm2c,
  wasmboxc} to convert it to multiple C source files, (2)
\texttt{libclang} to compile each generated C file into an optimized
x86-64 object file in parallel, and then (3) \texttt{liblld} to
combine all the generated object files into a single complete ELF
file.  We implemented this toolchain against the \sys API (upstreaming
our changes to \texttt{wasm2c}) and compiled it to Wasm.  It is
implemented as an ordinary \lang program, runs within \sys
normally, and is self-hosting: it can compile itself.

Machine codelets generated by this compilation toolchain are ELF files
containing relocation entries against the \sys API. \sys contains a small
in-memory ELF linker that links the codelet with the \sys API. This can be done
ahead-of-time and is not on the critical path.

\subsubsection{\sys API}

From the perspective of the original Wasm code, \sys's API allows it to
``map'' Blobs and Trees into native Wasm data types (for Blobs, a
read-only linear memory, and for Trees, an \texttt{externref}-typed table), and
create Blobs and Trees from native Wasm data types. This allows the
procedure to perform zero-copy conversions between \sys objects and native
structures.

Procedures interact with the \sys API through Handles represented as 256-bit
vector types, which can be passed by value as an \texttt{m256i} using AVX2
(\texttt{\%ymm}) registers on x86-64.  The allowed operations on a Handle are
decided by their type: BlobObjects and TreeObjects are mappable, making their data
accessible; Refs aren't mappable but procedures can inspect types and sizes of the referents;
Thunks can't be inspected at all, but new Encodes can refer to Thunks. By
creating Identification and Selection Thunks referring to Refs, and then
Encodes to those Thunks, procedures can request that \sys do the I/O
necessary to give a child function access to their contents.

\begin{small}
  \begin{lstlisting}[float,caption={\sys API},label=code1, style=cpp, basicstyle=\ttfamily\footnotesize, numbers=none]
// Map Fix data to native Wasm types
void attach_blob(m256i handle, wasm_memory*);
void attach_tree(m256i handle, wasm_table*);
m256i create_blob(int size, wasm_memory*);
m256i create_tree(int size, wasm_table*);
// Create Thunks
m256i application/identification/selection
  (m256i handle);
// Create Encodes
m256i strict/shallow(m256i handle);
// Query information about a Handle 
bool is_blob/tree/ref/thunk/encode(m256i handle);
int get_size(m256i handle);
\end{lstlisting}
\end{small}

\subsubsection{Security and Isolation}

\sys runs user-provided machine code in a shared address space. \sys ensures
isolation between function invocations:  Wasm procedures can only access their own
linear memory and make external function calls to \sys API, which provides
isolation similar to software-based fault isolation.
Procedures can also gain read-only access to \lang Blobs and Trees which
they get the Handles of by recursively mapping Trees, starting from the input to
\texttt{\_fix\_apply}. Proclets do not have
access to shared mutable memory or timers (both of which are nondeterministic),
which prevent timing side-channels. The safety and isolation provided by \lang
and \sys is similar to V8 isolates used by Cloudflare Workers platform.

\subsubsection{Adapting existing applications}

\begin{figure}[ht]
  \centering
  \includegraphics[width=0.8\columnwidth]{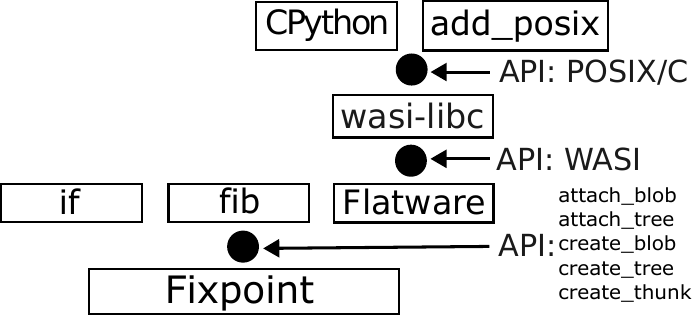}
  \caption{\lang can express ``native'' functions as well as Unix-style
  programs that manipulate a filesystem.}
  \label{fig:wasi}
\end{figure}

We have begun implementing a library to let existing Unix-style programs run on
\sys. The Wasm community has created a standard C library (wasi-libc)
that implements the C/POSIX interface in terms of underlying system calls known
as the Wasm System Interface (WASI). In turn, we implemented a library
called ``\posixlib'' (Fig.~\ref{fig:wasi}) that implements the WASI interface
in terms of the \sys API---treating the Thunk's arguments as containing a
Unix-like filesystem. \sys is oblivious to this translation layer; from its
perspective, it is an ordinary unprivileged part of the procedure.

\sys, via \posixlib, runs an off-the-shelf compilation of Python 3.12 built for
wasm32-wasi by VMWare Labs ~\cite{python3.12} with no modifications. The Python
script and arguments are passed in as part of the Thunk's definition, and the
\texttt{stdout} containing the result is returned as part of the output.

\subsection{Runtime architecture}

\begin{figure}[ht]
  \centering
  \includegraphics[width=\columnwidth]{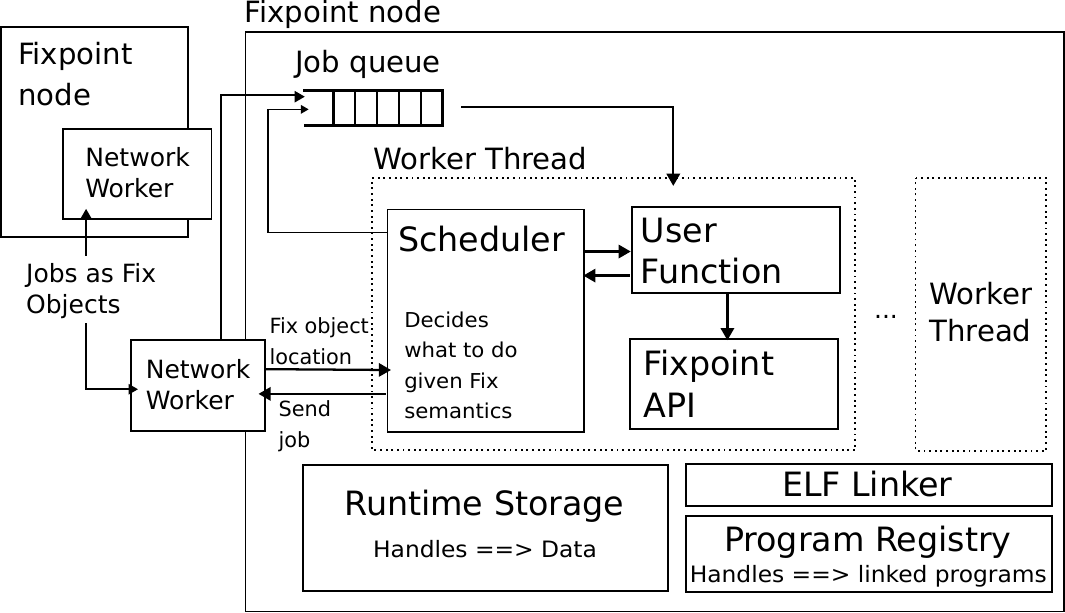}
  \caption{\sys infrastructure}
  \label{fig:sys-structure}
\end{figure}

\subsubsection{\sys workers}
The execution engine of \sys consists of a thread pool of \sys worker threads.
All worker threads share access to a queue of pending jobs and a runtime storage
that maps from Blobs and Trees to their data and from Encodes to evaluation
results. Each worker thread contains an instance of the Scheduler, which decides
what I/O and computations are needed for evaluating an \lang object based on
\lang semantics. \lang's computation model guarantees that a procedure cannot
block on further I/O before it returns, which allows the worker thread to directly jump to the
entry point of the codelet instead of launching new threads or processes.

 % \item \textbf{DependencyGraph. } Given a job, the DependencyGraph tracks the
 %   list of jobs that need to be completed to make progress on a job. When a
 %   worker finishes a job, it notifies the DependencyGraph of the completion,
 %   and pushes any unblocked jobs to the job queue.

%To apply a combination, a worker (1) verifies the procedure is created by the
%compilation toolchain, (2) invokes the in-memory ELF linker to link the codelet
%if it hasn't already been linked, (3) allocates per-execution data structures
%for the procedure, and (4) jumps to the entry point of the machine codelet.
%Although the current implementation executes step 2 and 3 immediately before
%applying the combination, we emphasize that these steps can be executed
%ahead-of-time, analogous to how current FaaS platforms anticipate future user
%workload and maintain a pool of warm instances.

\sys instances can interact to form a distributed execution engine. Each node
keeps track of which other nodes to notify when a job is completed, with a
Network Worker that handles messaging between nodes.

A \sys node delegates jobs to remote nodes by sending \lang values---Blobs and
Trees. This allows different nodes to communicate complicated user-function
dataflow without any intermediate conversion.  Moreover, as all dependencies
are specified as part of the packed binary format, \sys doesn't need to
maintain a global data structure or perform multiple roundtrips between nodes
to translate between a Tree and the function invocation it represents.

\subsubsection{Dataflow-aware distributed scheduling}

In the distributed execution engine, the Schedulers could also decide where a
job should be executed. In the current implementation, \sys
does not have a centralized scheduler, and each local scheduler makes
independent scheduling decisions by consulting a ``view'' of which data exist on
what node provided by the Network Worker.  When two \sys nodes first connect, they each provide the other
with a list of objects available locally, and this ``view'' is advanced
passively: if the other node sends or proposes to send an object, the object is
added to the view of the other node.  This ``view'' of where the data are
located will be replaced by a
distributed object store in the future.

Given an Application Thunk, the \sys scheduler descends the Application Thunk
Tree given \lang semantics, and collects its dependencies on data and the results
of other computations. For Encodes without results, it decides where to evaluate
the Thunk by picking the node that needs minimal data movement given locations
of existing \lang objects. Applications can ``hint'' an estimated
output size of a Thunk, and if so, the scheduler includes the cost of moving the
output as part of the data movement cost. In addition,
\lang allows the scheduler to see jobs that are dependees of the same
downstream job, and outsource parallel jobs to different nodes.

The choice of optimizing for data movement specifically is not intrinsic to \sys. With I/O
externalization, service providers have the flexibility to implement schedulers
that optimize for platform-specific goals, e.g. better bin-packing of RAM
resources.

%% file: evaluation.tex
\section{Evaluation}
\label{sec:eval}

In this section, we describe our experiments and performance measurements. We compared
\lang programs, running on \sys, with comparable applications written for a
serverless-computing system (OpenWhisk, MinIO, and Kubernetes), a
serverless-workflow-orchestration system (Pheromone), a serverless system with
Wasm runtime (Faasm), and a cluster-orchestration
system (Ray). Our evaluation answers five questions:

\begin{itemize}[noitemsep, topsep=0pt, itemsep=0pt]
  \item How do \lang and \sys improve function invocation and orchestration
    (Section \ref{invocation})?
  \item How does I/O externalization benefit resource utilization (Section
    \ref{externalization})?
  \item How does \lang's richer features allow computations to be broken down
    into much finer-grained (Section \ref{finegrained})?
  \item Can developers port real-word applications to \lang and gain
    performance benefit (Section \ref{compilation})?
  \item What is the experience of porting existing third-party software to
    \lang (Section \ref{sebs})?
\end{itemize}

\subsection{Benchmark Setup}

\textbf{Baselines } We compare \sys with 4 baselines:

\textbf{OpenWhisk. } OpenWhisk is a popular open-source serverless
platform. We deployed OpenWhisk on Kubernetes with MinIO, an
open-source object store. OpenWhisk is configured with Kubernetes as
the container factory, such that Kubernetes handles scheduling and
placing of OpenWhisk function invocations. By comparing with
OpenWhisk, we aimed to investigate whether \lang's approach of
externalizing I/O to the runtime produces a measurable benefit in
practice; in terms of allowing programs to achieve higher utilization
and decrease end-to-end execution time.

\textbf{Ray. } Ray is an open-source distributed execution framework with
    a with two main abstractions:
    \texttt{ObjectRef}s and \texttt{ray.get}. To get the values associated with
    a \texttt{ObjectRef}, users of Ray either 1) pass a \texttt{ObjectRef} as the argument to a
    Ray function invocation or 2) call \texttt{ray.get} on the \texttt{ObjectRef}
    which blocks the current function until the data are loaded.
    \texttt{ObjectRef}s and \texttt{ray.get}s are analogous to Thunks and Encodes of
    \lang.

    We compare \sys with three styles of Ray programs:

      \textit{\textbf{Ray + MinIO.} } User functions are Linux executables that
        reads from and writes to MinIO. The binaries of user functions locate on
        a single machine. When a function is invoked, Ray checks whether the
        binary is presenting locally, loads the binary if not, executes the
        binary via \texttt{Popen} and blocks until the subprocess returns. For Linux
        executables, this is the only viable style, as such user functions can not
        directly interact with Ray interfaces.

      \textit{\textbf{Ray (blocking-style I/O). }} User functions are Python functions
        implemented against Ray API. When a \texttt{ObjectRef} is needed, the
        function calls \texttt{ray.get} on the \texttt{ObjectRef}. This is how we
        expect Ray is normally used.

      \textit{\textbf{Ray (continuation-passing-style I/O). }} User functions are Python
        functions implemented against Ray API. Different from the previous
        usage, \texttt{ray.get} is never called. Whenever a \texttt{ObjectRef}
        is
        needed, a new function is invoked with the \texttt{ObjectRef} as the
        input,
        which breaks down applications into fine-grained function invocations
        along boundaries of I/O. This is the closest usage of Ray to \lang.

%    The major difference between the abstractions of Ray and \lang is that \lang
%    has abstractions for nested dependencies including 1) nested function
%    invocations (i.e. recursively forcing the Thunk until the result is a
%    Blob/Tree), 2) nested structures of references (i.e. Tree) and 3) precise
%    descriptions of dependencies on a subset of references from a Tree (i.e.
%    Selection and Identification Thunks).

    To illustrate how the latter two usages of Ray differ from each other and
    \lang, we describe how retrieving an entry from a linked-list is implemented
    in Listing ~\ref{lst:llrayb} and Listing ~\ref{lst:llraycps}.
    Ray (continuation-passing-style) has similar conceptual benefits to \lang: functions
    blocked on I/O don't occupy memory, are broken into fine-grained movable invocations,
    and allows Ray to utilize locality information.
    For these two usages of Ray, we setup Ray to read from the same file
    directory as \sys, and Ray has the same information as
    \sys in terms of where data locates across the nodes.

\begin{figure}[h]
    \begin{lstlisting}[language=Python, style=python, label={lst:llrayb},
    caption={In Ray (blocking-style), getting an entry is implemented as a Ray
    function that gets and blocks on the data of the next node until it
    reaches the node it needs. After the function invocation is started, data
    for Node objects are moved to where the function invocation is.}]
# A linked-list Node holds two Ray ObjectRefs: the data and the next Node
Node = Tuple[Ref, Ref]

@ray.remote
def get_blocking(head: Node, i: int):
  current_node = head
  for _ in range(0, i):
    # When data of the next Node is needed, the function blocks on the data
    current_node = ray.get(current_node[1])
  return ray.get(current_node[0])\end{lstlisting}
\end{figure}

\begin{figure}[h]
    \begin{lstlisting}[language=Python, style=Python, label={lst:llraycps},
    caption={In Ray (continuation-passing-style), there is no function
    blocked while Ray performs I/O to get the data, and Ray can choose different
    execution locations for each new function invocation.}]
@ray.remote
def get_cps(node: Node, i: int):
  if (n==0):
    return node[0]
  else:
    # When data of the next Node is needed, the function calls itself with the new data dependency, and the new function invocation is called when the data is ready
    return get_cps(node[1], i-1)\end{lstlisting}
\end{figure}

%begin{figure}
%    \begin{lstlisting}[language=C++, style=cpp, label={lst:llcoral},
%    caption={In \lang, the same function is implemented as creating a nested
%    Selection Thunk.}]
%// A linked-list Node is a Tree containing two Coral objects: the data and the next Node
%typedef m256i Node;
%m256i get(Node node, int i) {
%  auto ref = node;
%  for (int j = 0; j < i; j++ ) {
%    ref = create_selection_thunk(ref, 1);
%  }
%  return create_selection_thunk(ref, 0);
%}\end{lstlisting}
%\end{figure}

%    To get the 3rd entry of a linked list, Ray (blocking-style) needs one Ray
%    function invocation that blocks on getting new data three times; Ray
%    (continuation-passing) involves three individual Ray function invocations that do
%    not block during their invocations; \lang involves one \lang function
%    invocation that does not block to create the nested Selection Thunk. Difference
%    across Ray (blocking-style), Ray (continuation-passing) and \lang affects how the
%    systems handle the user tasks as shown in Figure \ref{fig:bptree}, which we
%    will discuss more in the following sections.

    By comparing with Ray, we would like to show how \lang's computation model
    expresses dataflow dependency at a lower overhead compares to previous
    distributed job execution system, and how \lang enables user procedures
    expressed as machine code to make their dataflow visible.

\textbf{Pheromone. } Pheromone allows users to describe dataflow of serverless
applications by specifying dependencies
between functions (e.g. invoke function B on the output of function A) or
dependency of a function on a set of data (e.g. invoke function A on any data
added to bucket B). Pheromone's approach targets reducing function orchestration
overhead by collocating intermediate data and function dependency information.
We deployed Pheromone on Kubernetes. By comparing with Pheromone, we would like
to show how \lang's computation model allows functions to specify dependencies
on both intermediate data and external data from durable storage and achieves collocation
of function data- and control flow with a more expressive dependency abstraction.

\textbf{Faasm. } Faasm is a serverless runtime that achieves lightweight
isolation by using Wasm for software fault isolation. Faasm's
isolation mechanism is similar to \sys, but without I/O externalization.
Therefore, Faasm provides functions with a host interface that supports
operations such as file system I/Os, shared states across multiple running
function invocations, etc., which leads to a more general interface than \sys,
but also a heavier runtime overhead.
We deploy Faasm locally using Docker containers. By comparing with Faasm,
we would like to show \lang's abstraction allows user programs to declare
precise code dependencies and let \sys minimize runtime overhead.

\textbf{Hardware } The experiments were run on \texttt{m5.8xlarge} Amazon EC2
instances. At the time of writing, an \texttt{m5.8xlarge} instance has
32 vCPU cores and 128 GiB memory. The volumes we used are Amazon EBS gp3 volumes
with 3,000 IOPS.

\subsection{Function invocation and orchestration} \label{invocation}

\subsubsection{Invocation overhead}
\begin{figure*}[ht]
    \centering
    \begin{subfigure}[b]{0.57\textwidth}
        \centering
        \includegraphics[height=1.22in]{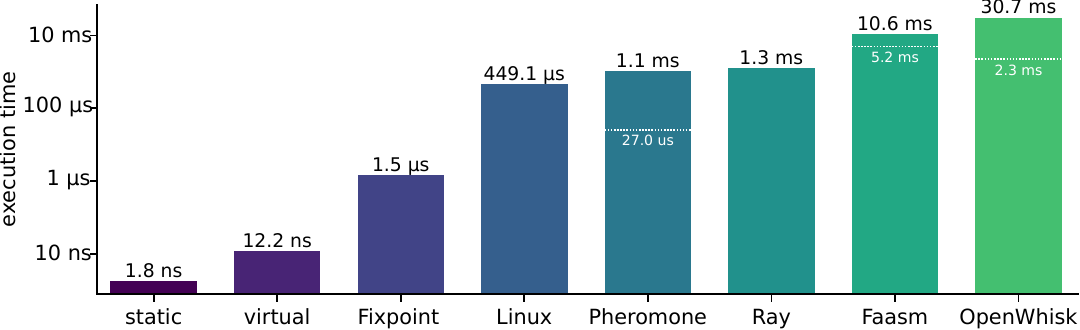}
        \caption{}
        \label{fig:lwvirt}
    \end{subfigure}%
    ~
    \begin{subfigure}[b]{0.4\textwidth}
        \centering
        \includegraphics[height=1.25in]{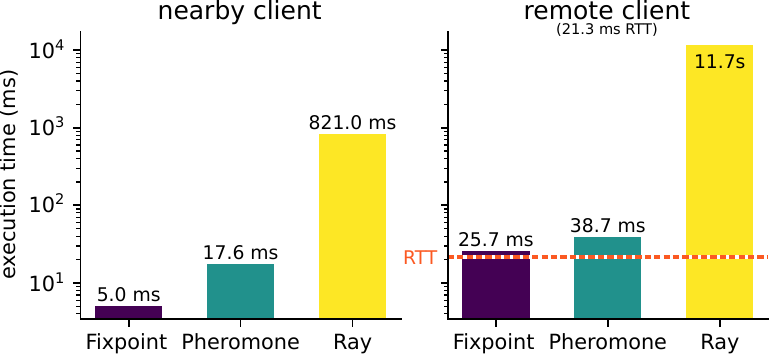}
        \caption{}
        \label{fig:functionchain}
    \end{subfigure}
    \caption{(a) Duration of a single trivial function invocation on
      \sys and comparator systems. \sys's overhead is lower, letting
      applications express themselves in finer-grained steps that each
      capture the minimum data needed. Dashed lines show ``core''
      execution time recorded by some systems' internal timers
      (generally excluding orchestration and cleanup). (b)~Duration of
      a chain of $500$ function invocations, with the client either
      near the server (left) or farther away (right). Ray incurs 500
      network roundtrips, while \sys and Pheromone can express the
      intended control flow in one go. For applications with pipelines
      of many fine-grained invocations, it's helpful to reduce the
      overhead of function composition as much as possible.}
\end{figure*}

In this section, we measure the raw overhead of \lang invocations by
measuring the performance of a trivial function\textemdash{}add two 8-bit numbers\textemdash{}run
in different types of isolation mechanisms. We compare the latency of executing
8 functions:

  \textbf{static}: Calling a statically linked function in C.

  \textbf{virtual}: The same, called as a virtual function in C++.

  \textbf{\sys}: The same, implemented in Wasm against the
    \lang API. This and the below approaches provide various forms of containerized or
    visible dataflow.

  \textbf{Faasm}: A C++ function against Faasm's API, pre-compiled.

  \textbf{Pheromone}: A C++ function against Pheromone's API.

  \textbf{Ray}: A Python function that adds two 8-bit integers, called as a
    Ray remote function.

  \textbf{OpenWhisk}: A full C/POSIX program that takes a JSON input, invoked as an OpenWhisk action.

  \textbf{Linux Process}: A full C/POSIX program that takes two 8-bit
    integers as command line arguments.

\textbf{Benchmark}: For the first seven, we evaluate the add function
4,096 times, and report the average time per function call. \sys, Pheromone, Ray
and OpenWhisk have different way of setting up a function. \sys statically links
the executable; Pheromone dynamically load the function as a shared library; Ray
pickles the Python function; OpenWhisk sets up the function container. To
exclude the function setup time from the measurement, we evaluate the add
function one time before the  measurement, and the time reported does not
include function
setup time. For the Linux process, we \texttt{vfork} the add program and
wait for its completion 4,096 times, and average the time per execution.
We report the average of five benchmark runs. For OpenWhisk, Pheromone and
Faasm, we also collect the core function logic execution time reported by the
systems, shown as stacked bars in the figure.

\textbf{Analysis}: As Fig.~\ref{fig:lwvirt} shows,
\texttt{add} as static or virtual function calls are the fastest
implementations, taking \qty{1.8}{\nano\second} and \qty{12.2}{\nano\second} for
execution, but do not provide isolation. Executing \texttt{add} as a Linux
process provides isolation, at the cost of a $>$\qty{400}{\micro\second}
context-switching penalty.

By contrast, the \sys program has an overhead of about
\qty{1.5}{\micro\second}. This overhead suggests a lower bound on the practical
granularity of an individual function invocation in \lang; in order to provide
$>50\%$ efficiency, each invocation will need do about \qty{1.5}{\micro\second}
of computation ($>$3 kcycles). This, in turn, suggests a minimum granularity of
data ``footprint'' that will be efficient in a multi-stage application.

Although \sys isn't close to the overhead of a function call, it is
much lighter than comparator systems: \sys is about $880\times$ lighter
than Ray and $20,000\times$ lighter than OpenWhisk. \sys is also $3500\times$
lighter than Faasm. \sys's speed-up is not solely attributable to its choice of
isolation mechanism, but also benefits from \lang's abstractions. These differences
in overhead suggest that \lang programs will be able to afford to
break down computation into much finer-grained and \emph{smaller}
containers.

\subsubsection{Orchestration overhead}

In this section, we measure the overhead of \lang function orchestration by
measuring the performance of a function chain.

\textbf{Benchmark. } Each function invocation increments its input value by 1,
which is consumed by the next function invocation. We measure the performance
of 500 chained functions, as shown in Fig.~\ref{fig:functionchain}. We run the
experiment with the clients placed on one of the machine in the EC2 cluster and
on a remote server and report the average latency over 5 runs. We invoke the
function once before taking the measurements, such that the measured results do
not include the time of function loading for all three systems.

\textbf{Analysis. } Ray allows users to specify dependency on the level of
individual function invocations, but the specified dependencies are coupled with
the location where they are specified. Ray needs to pay for a roundtrip to the
client for each dependency resolution.

Compared with Ray, Pheromone restricts the expressivity of its dependency model to
the level of individual functions. It decreases the overhead of
sharing function dependencies in a distributed setting, and enables Pheromone to
collocate function dependency information and function outputs. Pheromone is
$47\times$ faster and $303\times$ faster than Ray with different client
locations.

\lang allows users to specify dependency on the level of
individual function invocations. This gives \lang the
level of expressivity of Ray and Pheromone's capabilities of
coupling function data- and control flow.

\subsection{I/O externalization} \label{externalization}
\lang has a computation model that externalizes all I/Os of user function
invocations, making it possible for service providers to late binding
physical resources and optimize placement decisions with locality information.
In this section, we focus on how I/O externalization benefits 1) one-off functions
that depend on input from network storage and 2) a map-reducing style workload
with locality information.

To better understand the contributions of \lang and \sys, we conduct the
experiments with two ablations:

\newcommand{\nolocality}{\sys (no locality)\xspace}
\newcommand{\internalio}{\sys (``internal'' I/O)\xspace}
\newcommand{\nolocalityandio}{\sys (no locality + ``internal'' I/O)\xspace}

\textbf{\nolocality}: The scheduler picks a random execution location
for each function invocation.

\textbf{\internalio}: Function invocations occupy the claimed
physical resources before their I/O dependencies are resolved. In addition, we
oversubscribe the number of CPU cores and do not oversubscribe the memory for
\sys, which is analogous to how status quo serverless platforms manage physical
resources.

\subsubsection{I/O externalization for one-off functions}
We measure the benefits of \lang's I/O externalization for
individual function invocations that depend on inputs residing on
a remote data server configured with 150ms response latency to mimic ~Amazon~S3 performance
of fetching small objects~\cite{s3perf}.

Each function invocation reads an input that resides on the remote data server,
and adds the input to itself. Different function invocations depend on different
inputs. Each function invocation requests 1 CPU
and 1 GB memory, and the \sys server is configured with 32 available cores and
64 GiB available memory, which allows up to 32 function invocations that \sys
has fetched the inputs for to run in parallel. In \internalio, we oversubscribed
the CPU cores to 200, which allows up to 64 function invocations to fetch the
inputs and calculate the result in parallel.  This is analogous to current serverless services
where a function invocation starts fetching data from
network storage after it is placed and the requested physical resources are
provisioned. We run 1,024 function invocations for each run, recorded the CPU
utilization data on the server and report the
average latency of 5 runs in Fig.~\ref{fig:ioextern}.

\textbf{Analysis. } I/O externalization allows \sys to achieve late binding of
physical resources: resources are allocated after the data for function
invocations is ready. As a result \sys is $8.7\times$ faster than \internalio. This suggests that typical serverless
workloads---individual function invocations that depend on input from network
storage---could benefit from \lang's computation model.

\definecolor{good}{rgb}{0.85,1.0,0.92}
\definecolor{bad}{rgb}{1.0,0.93,0.93}

\newcommand{\good}{\cellcolor{good}}
\newcommand{\bad}{\cellcolor{bad}}

\subsubsection{I/O externalization with locality information}
\begin{figure*}
  \begin{subfigure}[b]{0.33\textwidth}
  \centering
  \footnotesize
  \begin{tabularx}{2.13 in}{l r r}
    & \lang & \lang \\
    & & \raisebox{4 mm}{\tiny (+ ``internal'' I/O)}\\[-0.4 cm]
    \hline
    user & \qty{3}{\milli\second} & \qty{11}{\milli\second}\\
    system & \qty{2}{\milli\second} & \qty{6}{\milli\second} \\
    I/O + wait & \qty{263}{\milli\second} & \bad \qty{2621}{\milli\second} \\
    \hline
    \bf total & \qty{268}{\milli\second} & \bad \qty{2638}{\milli\second}\\
    \hline
    \bf throughput & \qty{3827}{tasks\per\second} & \bad \qty{388}{tasks\per\second}\\
  \end{tabularx}

  \vspace{2 \baselineskip}

    \caption{}
  \label{fig:ioextern}
  \end{subfigure}
  ~
  \begin{subfigure}[b]{0.65\textwidth}
  \includegraphics[height=1.32in]{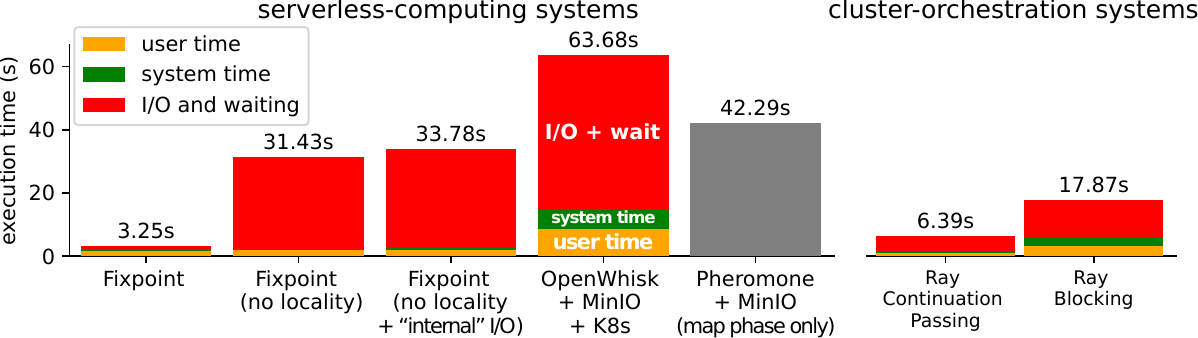}
  \caption{}
  \label{fig:wikipedia}
  \end{subfigure}
  \caption{(a) Duration to execute 1,024 function-invocation
    requests on a server with 32 CPUs and 64 GiB RAM. Each invocation
    depends on a single input from a network storage with a
    \qty{150}{\milli\second} response time and asks for 1~CPU core and
    1~GB memory. I/O externalization allows \sys to fetch data
    dependencies before committing physical resources, greatly
    increasing task throughput. (b)~Counting occurrences of a
    3-character chunk in 984 100-MiB shards from Wikipedia.  \sys's
    runtime is much lower than other systems'---fueled, in large part,
    by its ability to avoid stalling CPUs to wait for dependencies.}
\end{figure*}

  We count the occurrences of a 3-character string through
    a dump of the English Wikipedia sharded into 984 100 MiB chunks in a
    map-reducing style.
      The workload consists of 2 functions: (1) \texttt{count-string}
      takes a chunk and a string as inputs and reports the number of occurrences
      of the string and (2) \texttt{merge-counts} merges the results in a
      binary reduction. \texttt{count-string} is invoked on every Wikipedia chunk
      and \texttt{merge-counts} is invoked on the results of every two completed
      function invocations until the final result.
    We deploy \sys, OpenWhisk and Ray on a 10-node cluster with 320 vCPUs in
    total. The 100 MiB
    chunks are scattered among the 10 nodes randomly for \sys and Ray, and
    store in MinIO deployed on the same cluster for OpenWhisk. We measure the
    end-to-end execution time of \sys, \nolocality, \nolocalityandio, Ray
    (continuation-passing), Ray (blocking-style) and OpenWhisk, and record CPU
    utilization data of the 10 nodes from \texttt{/proc/stat}. Results are shown
    in Fig.~\ref{fig:wikipedia}, averaged over 16 runs.

    \nolocality selects a random node
    for each invocation of \texttt{count-string}. On top of that, \nolocalityandio oversubscribes the CPU, running 128 threads instead of 31.

    Due to the implementation of Pheromone, we are not able to get the reduce-phase
    (i.e. \texttt{merge-counts}) to run: before map-phase function invocations
    complete, Pheromone starts the reduce-phase function invocations which try
    to access the output bucket of the map phase with a hard-coded timeout value.
    Pheromone's implementation makes it hard to precisely record the CPU
    utilization data, as its components have 100\% CPU utilization whether there
    are user applications running or not. We report the end-to-end execution time of
    the map phase in Pheromone, averaged over 5 runs.

\textbf{Analysis. } I/O externalization allows \sys to pick execution locations
of functions with the knowledge of data locality information, and \sys achieves
$9.7\times$ speed-up compared to \nolocality. Ray sees
similar benefits as Ray (continuation-passing-style) breaks the program into
fine-grained invocations such that Ray can pick the execution location before
the function is started.

\nolocalityandio
oversubscribes CPUs such that no \texttt{count-string} invocation is blocked due
to physical resource limitation, but the oversubscription introduces a $7.5\%$
overhead.

Compared to \lang, Pheromone's dependency abstraction does not allow users to
specify function dependencies on data that are not intermediate results.
Although the end-to-end execution time of Pheromone does not reflect the actual
performance of a production-grade system implemented with similar
ideas, its execution time is of a magnitude that is more analogous to \nolocality.

\subsection{Fine-grained function invocations} \label{finegrained}
\begin{figure}
  \centering
  \includegraphics[width=0.98\columnwidth]{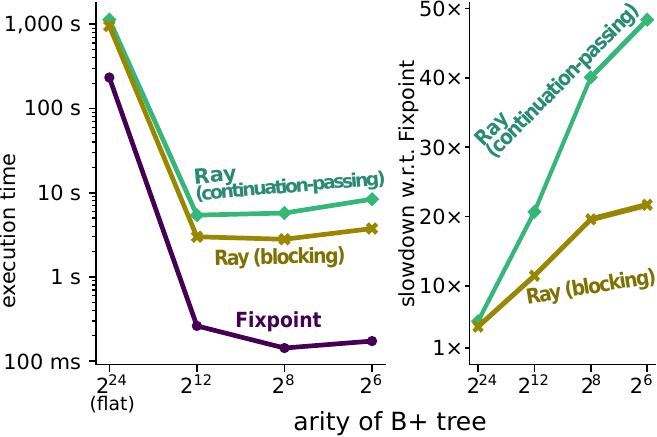}
  \caption{Applications with data-dependent dependencies benefit from
    \lang's model. Plotted here is the time to traverse a 6M-entry
    B+-tree of Wikipedia article titles, searching for one entry. Each
    step of the search examines one node to decide which node to
    descend into. A Ray program (broken into fine-grained invocations
    in continuation-passing style) scales poorly as the tree arity
    decreases. A blocking Ray program performs better here, but worse
    in Fig.~\ref{fig:wikipedia}. Compared with \lang, both Ray
    implementations are hindered by a lack of expressiveness regarding
    data and control flow.}
      \label{fig:bptree}
\end{figure}

\begin{table*}[ht]
\centering
\small
\begin{tabular}{ |c|c|c|c|c|c|c| }
  \hline
   Tree arity ($a$) / Tree depth ($d$) & $2^{24} / 1$ & $2^{12} / 2$ & $2^{10} / 3$ & $2^{6}/ 5$ & \multirow{2}{*}{data
   accessed} & \multirow{2}{*}{maximum memory footprint} \\
   \cline{1-5}
   & \multicolumn{4}{c|}{\# of function invocations} & & \\
   \hline
   \sys & \multicolumn{4}{c|}{$d$} & $ad$O(key size) & $a$O(key size) \\
   \hline
   \shortstack{Ray (Continuation Passing)} & \multicolumn{4}{c|}{$2d$}
   & $ad$(O(key size) + O(entry size)) & $a$(O(key size)+O(entry size)) \\
   \hline
   Ray (Blocking) & \multicolumn{4}{c|}{1} & $ad$(O(key size) + O(entry size))
   & $ad$(O(key size) + O(entry size)) \\
   \hline
\end{tabular}
\caption{The amount of data accessed and maximum memory footprint of \sys
  and comparator systems to get a value from B+ Trees of different arities that
  hold all Wikipedia article titles. O(key size) stands for the length
  of Wikipedia article titles and O(entry size) stands for the size of Tree entry.}
\label{table:bptree}
\end{table*}

Compared to status quo serverless workloads\textemdash{}individual function
invocations\textemdash{}other real-world applications have more dynamic data
dependencies. They may traverse tree-like structures and require several rounds
of I/O, for example, identifying objects given a BVH tree of scene data for 3D
rendering, or fetching a file from a file system represented by nested trees.
Breaking down these applications into fine-grained function invocations reduces
their memory footprint, but may hurt overall performance due to the overhead of
function invocation and orchestration. In this section, we would like to show
how \lang and \sys allow users to describe data dependencies precisely and
benefit from breaking applications into fine-grained steps.

\textbf{Benchmark. } We take the list of titles of English Wikipedia articles,
about 6 million entries with an average length of 22 bytes, and create B+ Trees
of different arities with the titles as keys (details shown in Table \ref{table:bptree}). Each node of
the tree is implemented as a list of \texttt{ObjectRef}s for Ray and as a Tree
of for \sys. An internal node contains the \texttt{ObjectRef}/Handles to
subtrees. A leaf node contains the \texttt{ObjectRef}/Handles to values. All
nodes contain the \texttt{ObjectRef}/Handle to an array of keys. Traversing the
keyspace of the B+ Trees descends the B+ Tree node-by-node.  For each layer,
Ray gets 2 \texttt{ObjectRef}s: one for a list of child \texttt{ObjectRef}s,
and one for the array of keys corresponding to each child. For each
\texttt{ObjectRef}, Ray (blocking-style) does a blocking \texttt{get} and Ray
(continuation-passing-style) makes a new Ray function call.

We deploy \sys and Ray on a single node, and configure \sys and Ray to each
use a single worker thread. All the data of the B+ tree is stored on the same
node for \sys and Ray.
We run five independent sets of queries, resetting the system state between each set.
Each set of queries contains 10 sequential queries for different
keys chosen randomly. We measure the end-to-end execution time
of executing a set of queries, and average the results across the five set
of queries. We run this experiments twice for arity $2^{24}$, and 16 times
for all other arities, results in Fig.~\ref{fig:bptree}.

\textbf{Analysis. } As the arity of B+ Tree decreases, the total data accessed
and maximum memory footprint of the 3 comparators decrease, with a different
factor across different implementations at the cost of increased number of
function calls.

Although Ray (continuation-passing-style) outperforms Ray (blocking-style) in
Fig.~\ref{fig:wikipedia} as it shares similar benefits as \lang, Ray
(continuation-passing-style) involves more function invocations. As the function
invocations get more fine-grained in this benchmark, the extra function
invocation and orchestration overhead makes it a net loss for Ray
(continuation-passing-style), which consistently performs
worse than Ray (blocking-style) and sees an increase in end-to-end execution time as
the arity decreases from $2^{12}$ to $2^{8}$, while \sys sees a decrease in
end-to-end execution time. At the arity of $2^{6}$, Ray (blocking-style) is
$22.3\times$ slower; and Ray (continuation-passing-style) is $49.9\times$ slower.

This suggests that \lang's richer semantics and lower overhead of function
orchestration makes it possible for user and the platforms to benefit from smaller
memory footprint and amount of data accessed by decomposing applications into
multiple invocations of fine-grained data dependency, which could benefit
applications with dynamic data dependency mentioned before.

\subsection{Burst-parallel application} \label{compilation}
\begin{figure}
  \centering
  \includegraphics[width=0.7\columnwidth]{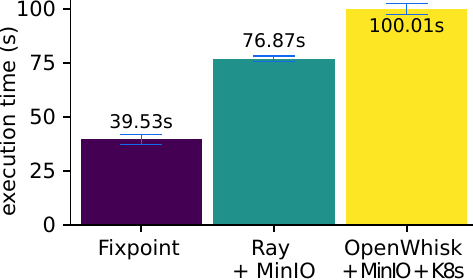}
  \caption{Duration to run a large software-compilation job, with a
    dataflow graph that evolves over time, on a 10-node cluster with
    320 total vCPUs. \lang lets \sys outsource function invocations
    with more fine-grained data needs, and bundles the
    data-dependencies along with the invocations. Error bars represent
    standard deviation of 5 runs.}
  \label{fig:compilation}
\end{figure}

In this section, we would like to study whether users can port real-world
applications to \lang and see performance gain. We port \texttt{libclang} and
\texttt{liblld} to \lang, which involves three parts of programmer efforts: we
made a one-line change to upstream LLVM codebase to remove its dependency on
threading support; we created function stubs for functions that wasi-libc does
not implement; we wrote driver programs that interact with \lang for I/Os,
which includes 186 lines of code in C, and 93 lines of code in WAT (the POSIX
counterparts are implemented with 92 lines of code in C).

\textbf{Benchmark. } To measure \sys's performance on burst-parallel
applications, we compile a project with almost 2,000 C source files, resulting
in parallel invocations of \texttt{libclang} (each depends on a input C file,
plus system and clang headers) and a single invocation of \texttt{liblld} to
combine the outputs into a single object file.

For OpenWhisk, \texttt{libclang} and \texttt{liblld} are created with
Docker images due to OpenWhisk's limit on binary size. Other data
dependencies and input/output files are stored in MinIO. OpenWhisk functions
are created when needed, and the reported execution time includes function
creations.

For Ray, each executable behaves the same as in OpenWhisk, and Ray launches
executables via \texttt{Popen}. The executables start on a single node. When a
Ray job is scheduled on a node, it first checks whether the executable exists
on the machine, and loads the executable if not. As Ray does not provide
interfaces of getting data for such executables, the executables read data
dependencies from MinIO, similar to OpenWhisk.

For \sys, all dependencies including data and binaries are uploaded from the
client at execution time. The \sys client is connected to a single \sys server
node.

We deployed \sys, OpenWhisk and Ray on a 10-node cluster with 320 vCPUs in
total, and measured the end-to-end execution time of \sys, Ray + MinIO, and
OpenWhisk. Results are shown in fig.~\ref{fig:compilation}, averaged over 5 runs.

\textbf{Analysis. } \lang allows user programs expressed in machine code, like
    \texttt{libclang} and \texttt{liblld}, to make their dataflow visible to
    \sys in a language-agnostic way. \sys also has a lower function invocation
    and orchestration overhead, such that the large number of parallel function
    invocations are distributed across the nodes efficiently.
    \sys achieves a $1.9\times$ speed-up compared to Ray and $2.5\times$
    speed-up compared to OpenWhisk since Ray and OpenWhisk do not have the same
    level of visibility as \sys.

    %Porting \texttt{libclang} and \texttt{liblld} is straightforward, since
    %\texttt{LLVM} already separates its application logic into I/O cnad compute.

    %has a well-engineered build interface that excludes code paths
    %not well-supported by existing WASM toolchains, like threading. Given this
    %experience, we expect the difficulties of porting existing applications are
    %related to whether the application has a clear boundary between computation
    %logic and I/O logic (e.g. reading files, establishing network connections).

\subsection{Porting existing third-party software} \label{sebs}

In this section, we would like to show the user efforts involved in porting
existing applications with the support of \posixlib. We take two functions\textemdash{}
\texttt{dynamic-html} and \texttt{compression}\textemdash{}from SeBS~\cite{sebs}, a
popular serverless benchmark suite. \texttt{dynamic-html} takes a user name as
input, and generates an HTML from a template using \texttt{Jinja} library.
\texttt{com\-pression} takes a bucket name as input, downloads
all the files and creates an archive. Porting these two
functions involves two parts: (1) we modify the functions to read inputs from
command line arguments, and the data dependencies from the file system; and (2)
we identify dependencies of the functions. \texttt{dynamic-html} depends on external libraries and the HTML 
template, while \texttt{compression} depends on the files to create an archive for. 
We create \lang objects that represent the dependencies as files in a
Unix-like filesystem in the format required by \posixlib.
With these changes, we are able to run the two functions with an off-the-shelf
compilation of Python 3.12 built for wasm-wasi in \sys via \posixlib.

\begin{figure}[h]
\centering
  \includegraphics[width=0.95\columnwidth]{"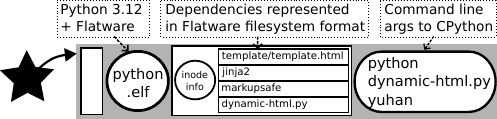"}
  \caption{Represent an invocation of \texttt{dynamic-html} with \lang objects.}
\end{figure}

These two parts are representative of the process of porting arbitrary
programs to \lang: (1) modifying the source code to remove code dependencies that are not yet
supported by WASI or \lang (e.g. socket operations, threads, etc.); and (2) identifying the minimum repositories that the programs will need access
to. Complexity of the porting process varies across different
programs. Applications that are mainly ``computational'' are easier to port
to \lang, while ones that heavily depend on nondeterministic services need either
refactoring the applications, or expanding \lang. Identifying the
minimum repository relies on application-specific knowledge. For applications
like serverless functions with well-defined dependencies, programmers
could include everything in the minimum repository, as what we did for the
two SeBS functions. For applications with highly-dynamic dataflow, how
they should be broken down into smaller invocations with evolving minimum
repositories relies on programmers' discretion.

%% file: limits.tex
\section{Limitations and Future Work}
\label{sec:limits}

\lang is a constrained model of computation, and this comes with
significant limitations.  Although some of these might be lifted with future
work, doing so will require solving some open research problems.

\textbf{\lang can only model functional workloads. } Although \sys shows
substantial performance benefits in functional workloads, \lang's model of computation
is more limited than existing FaaS
platforms. \lang requires user programs to be pure functions that
consume content-addressed data (or the outputs of other computations). We think this
model is compatible with many typical uses of FaaS platforms, where functions
use provider APIs to perform limited I/O to other services, but excludes
applications that rely on the ability to perform arbitrary network requests.
It also excludes applications that wouldn't fit the serverless model because of
a reliance on shared mutable state, e.g., multi-user
databases, message buses, e-commerce workloads~\cite{deathstar}, etc. Existing
applications that rely on such state are deployed on a combination of services:
computations are run on serverless services, while the shared mutable state is
managed by other external services like databases or blob storage. We imagine that \lang's role
in such applications will one day be similar to current serverless
platforms, but currently \lang doesn't support them.

\textbf{Nondeterministic I/O must be delineated. } \lang requires inputs
and dependencies of a function invocation to be identifiable deterministically.
However, real-world programs sometimes desire nondeterministic I/O, e.g., to
gather a random seed, the time, or arbitrary information from a remote
sensor or service. In \lang's world, this nondeterministic I/O needs to be
delineated from the rest of the program, so it can be performed externally by a
\lang runtime. We believe that it will be possible to statically or dynamically
transform existing programs with embedded I/O into a sequence of pure functions
with delineated I/O, using techniques similar to asynchronous programming
languages~\cite{asyncify}.

\textbf{I/O externalization is burdensome. }
Existing programs have to be recompiled with a \lang-targeting toolchain.
Programmers need to separate applications into I/O
and compute, so each stage of a computation can declare its dependencies
before execution, and convert existing programs to continuation-passing style to keep necessary state across points of I/Os. This
burden could be lifted by expanding \posixlib and providing
implementations of common programming paradigms, e.g. map-reduce, on
\lang. In addition, \lang's visibility into data- and control flow suggests the
possibility of lightweight continuation capture, where existing
programs are automatically split at I/O operations. We leave
implementation of this transformation to future work.

%\textbf{Performance penalty may outweigh the benefits for certain
%  applications. } Although the choice of WebAssembly is not intrinsic
%to \lang, compiling code through \sys's current compilation toolchain
%imposes slowdowns that have been benchmarked at upwards of
%14\%~\cite{wasmboxc}. There will certainly be applications, especially ones
%that do lots of single-threaded computation and little I/O, that
%experience more overhead compared to the benefits gained from \lang.

\textbf{Towards computation-as-a-service.} We are optimistic that changing the interface to computation on FaaS
platforms could make it possible to realize new efficiencies, to the
mutual benefit of operators and their customers.  We expect serverless platforms might be able to change dramatically:

\textit{Ultra-high-density multitenancy.} Serverless platforms will be
    able to pack as many applications as possible into limited memory and CPU
    resources, with fine-grained understanding of each application's time-varying memory
    footprint.

    \textit{Computational ``garbage'' collection.} Because \lang
    computations are deterministic products of known dependencies,
    users who opt for ``delayed-availability'' storage would grant the
    provider the ability to delete stored objects as long as the
    provider knows how to recompute them on demand,
    within the SLA window for the data to be delivered.

    \textit{``Paying for results.''} Billing models that reward better
    placement and scheduling strategies could benefit both users and
    providers. Operators could compute prices based on an ``upfront''
    cost (the size of an invocation's data inputs and RAM
    reservation), plus a ``runtime'' cost that immunizes the customer
    from bad placement or a ``noisy neighbor.''  E.g., instead of
    milliseconds of wall-clock time, the runtime cost might be a
    combination of ``instructions retired'' plus a penalty for L1 and
    L2 cache misses (which are the core's fault), but not L3 cache
    misses, which may be affected by neighbors on a CPU. Invocations
    that come with more-distant deadlines could carry a lower cost and
    allow a provider to spread out load.

\textit{Commoditizing cloud computing.} Because computations will have
a single, unambiguous result, providers could sign statements with
their answers---``$f(x)\rightarrow y$, according to Provider Z''---and
customers could bid out jobs to any provider that carries acceptable
``wrong answer'' insurance and double-check answers if and when
they choose.

%% file: concl.tex
\section{Conclusion}

In this paper, we presented \lang, an architecture for serverless computing
where functions and the underlying platform share a common
representation of a computation. This leads to better placement and scheduling of user jobs,
improving performance and reducing waste. \lang is available at \url{https://github.com/fix-project/fix},
with a Zenodo snapshot at \url{https://zenodo.org/records/17154970}.

%% file: acks.tex
\begin{acks}

We thank our shepherd Anil Madhavapeddy and all our reviewers for
their thoughtful feedback. We are grateful to Angela Montemayor, Amit
Levy, Mark Handley, Gerald Sussman, Sam Clegg, Deian Stefan, Sadjad Fouladi,
 Justin Weiler, Colin Drewes, Katie Gioioso,
Zachary Yedidia, Matthew Sotoudeh, Rupanshu Soi, Alex Aiken, Philip Levis,
Scott Shenker, Sebastian Thrun, and Alex Crichton for helpful conversations. This work was supported
in part by NSF grants 2045714 and 2039070, DARPA contract
HR001120C0107, a Sloan Research Fellowship, and by Google, Huawei,
VMware, Dropbox, Amazon, and Meta Platforms.

\end{acks}